\documentstyle[12pt]{article}
\newcommand{\be}{\begin{equation}}
\newcommand{\ee}{\end{equation}}
\newcommand{\bdm}{\begin{displaymath}}
\newcommand{\edm}{\end{displaymath}}
\newcommand{\bea}{\begin{eqnarray}}
\newcommand{\eea}{\end{eqnarray}}
\begin{document}
\title{{\hfill \tt to appear in Theor. Math.
Phys}\\{\hfill}\\{\hfill}\\Kovalevskaya top -- an elementary approach}
\author{A. M. Perelomov
 \footnote{On leave of absence from the
Institute for Theoretical and Experimental Physics, 117259, Moscow, Russia.
Current E-mail address: perelomo@dftuz.unizar.es}\\
{\small\em Departamento de F\'{\i}sica, Facultad de Ciencias,
Universidad de Oviedo,}\\{\small\em E-33007 Oviedo, Spain}}
\date{}
\maketitle

\begin{center}
To the memory of J{\"u}rgen Moser
\end{center}

\begin{abstract}
\noindent The goal of this note is to give an elementary and
very short solution to equations of motion for the Kovalevskaya top [1].
For this, we use some results from the original papers by Kovalevskaya [1],
K\"otter [2] and Weber [3] and also the Lax representation from the note [4].
\end{abstract}

\noindent
{\bf 1.} The Kovalevskaya top [1] is one of the most beautiful examples of
integrable systems. This is the top for which the principal momenta
of inertia $J_1$, $J_2$, $J_3$ satisfy the relation
\begin{equation}
J_1=J_2=2J_3=J, \end{equation}
and the center of mass lies in the equatorial plane of the body (for the
simplicity, we put further $J=1$). The dynamical variables
are components $m_1$, $m_2$, $m_3$ of angular momentum and components $n _1$,
$n _2$, $n _3$ of the center mass vector in the system related to the
principal axes of the body.

This system is Hamiltonian relative to the Poisson structure
for the Lie algebra $e(3)$ of motion of three-dimensional Euclidean space
\begin{equation}
\{m_i,m_j\}=\varepsilon _{ijk}\,m_k,\qquad
\{m_i,n_j\}=\varepsilon _{ijk}\,n_k,\qquad \{n_i,n_j\}=0,
\end{equation}
where $\varepsilon _{ijk}$ is a standard totally skew-symmetric tensor.

The Hamiltonian has the form
\begin{equation}
H=\frac{1}{2}\left( m_1^2 + m_2^2 + 2 m_3^2 - n_1\right)
\end{equation}
and the equations of motion are (the dot means a derivative in time)
\begin{equation}
\dot m _j =\{H,m _j\}, \qquad \dot n _j =\{H,n _j\} ,
\end{equation}
or in the explicit form,
\begin{equation} \begin{array}{lll}
\dot m_1=m_2m_3,&  2\,\dot m_2=-\,(2\,m_3m_1+n_3),& 2\,\dot m_3=n_2, \\
\dot n_1=2\,m_3n_2-m_2n_3,& \dot n_2=m_1n_3-2\,m_3n_1,&
\dot n_3=m_2n_1-m_1n_2. \end{array} \end{equation}
Note that the angular velocity vector has the form
\begin{equation}
(p , q, r )=(m_1 , m_2 , 2m_3 ) .\end{equation}

In the celebrated paper [1], Kovalevskaya succeeded in integration of these
equations in terms of abelian functions of two variables.
The Kovalevskaya approach was simplified later by K\"otter [2].
Note also the paper by Kolosov [5], where he reduced this problem to the
problem of motion of the point on the plane in a potential field.

One century was gone, and the Kovalevskaya top roused interest again.
In the paper [4] the Kovalevskaya top was considered as the projection
of the Euler top.
This approach gives as the explanation of famous relation (1) as
the natural multi-dimensional integrable generalizations of such system.

In papers by Enolsky [6], [7], nontrivial reductions were found which give
the elliptic solutions for the Kovalevskaya top. In the paper by Novikov
and Veselov [8] (see also [9] and references there), the action-angle
variables for this problem were constructed and the Poisson commutativity of
variables $s_1$ and $s_2$ was discovered. Note that namely these variables
are appeared at  the consideration of the Kovalevskaya top as the projection
of Euler's top.

Authors of number of papers (see [10], [11], [12], [13] and
references therein) used the algebro-geometrical approach to this
problem. Unfortunately, this approach is very complicated and till
now only some part of original results for the Kovalevskaya top
was reproduced in framework of it.

For example,  the Lax representation with spectral parameter [10],
[12] gives the spectral curve of genus three and correspondingly
the abelian functions of three variables, but not abelian
functions of two variables as in the original Kovalevskaya paper
[1]. Even for the simplest case $(m,n)=0$, the correspondence
between two such approaches is very complicated [13]. So, in
author's opinion, the original Kovalevskaya--K\"otter approach
being elementary and natural one is more adequate to the problem
under consideration.

\bigskip \noindent
{\bf 2.} Following [1] and [2], let us remind first that equations (5) have
four integrals of motion
\begin{eqnarray}
H_1&=&2\,H= m_1^2+m_2^2 +2\,m_3^2-n_1 = h_1,\\
{\tilde H}_2&=&\xi _+\xi _- = k^2, \\
C_3&=&(m,n)=m_1n_1+m_2n_2+m_3n_3 = c_3,\\
C_4&=& n_1^2+ n_2^2+ n_3^2 = c_4,
\end{eqnarray}
where
\begin{equation}
\xi _{\pm }=m_{\pm}^2 + n_{\pm},\qquad m_{\pm} = (m_1 \pm i
m_2),\qquad n_{\pm}=(n_1 \pm i n_2) . \end{equation}

Note that $C_3$ and $C_4$ are Casimir functions and the equations $C_3=c_3$,
$C_4=c_4$  define the four-dimensional symplectic manifold ${\cal M}_c$ --
the orbit of coadjoint representation of Lie group $E(3)$
(the group of motion of three-dimensional Euclidean space).

The integration of equations (5) consists from several steps.

\bigskip\noindent
{\bf 3.} We start with the Lax representation [4]
describing the Kovalevskaya top as the projection of the Euler top
\begin{equation} \begin{array}{l}
\dot L_2=[L_2,M_2],\\
L_2=-A\left( 2\,\hat{m}^2+(\gamma \otimes n+n\otimes \gamma )\right) A,\\
M_2=-A\,\hat{m}\,A, \end{array} \end{equation} where
\begin{equation}
\hat{m}=\left( \begin{array}{rrr} 0&m_3&-m_2\\ -m_3&0&m_1 \\
m_2&-m_1&0
\end{array} \right) ,\quad
A=\left( \begin{array}{ccc} 1&0&0\\ 0&1&0\\ 0&0&0 \end{array} \right) ,\quad
\gamma =(1,0,0),\quad n=(n_1,n_2,n_3),
\end{equation}
\begin{equation}
\mbox{tr}\,(L_2)=2H_1=4H,\qquad \mbox{det}\,L_2=H_2=H_1^2-{\tilde H}_2.
\end{equation}
Then we have
\begin{eqnarray}
&&\mbox{det}\,(sI-L_2)=s\,P_2(s),\qquad P_2(s)=s^2-2\,H_1s+H_2,\nonumber \\
&&H_1=h_1,\quad H_2=h_2,\quad h_2=h_1^2-k^2.
\end{eqnarray}

In the Kovalevskaya case, the equations of motion contain as
quadratic as linear terms in dynamical variables $m_1$, $m_2$,
$m_3$; $n_1$, $n_2$, $n_3$. From this it follows that some of
these variables being meromorphic functions of time $t$ have the
second order poles in $t$.

Let us try to find the change of variables such that the equations
of motion will contain only quadratic terms. This may be achieved
by elimination of variables $n_1$ and $n_2$.

From equations (7)--(10) follows two linear equations for
variables $n_1$ and $n_2$
\begin{equation} \begin{array}{l}
m_1n_1+m_2n_2=c_3-m_3n_3,\\
\left( m_1^2-m_2^2\right) n_1+2\,m_1m_2n_2=\frac12\left(
n_3^2-\left( m_1^2+ m_2^2\right) ^2+k^2-c_4\right) . \end{array}
\end{equation}
Using them we may eliminate $n_1$ and $n_2$ from equations
\begin{equation} 2\,H_1=2\,h_1,\qquad H_2=h_2. \end{equation}
Then we discover that the left hand side of these equations
becomes the quadratic form if we introduce new variables
\footnote{Note that these variables were used already in papers
[1] and [2]}
$f =(f_1,f_2,f_3)$, $g =(g_1,g_2,g_3)$,
where
\begin{eqnarray}
f_1&=&\frac1{m_2},\qquad f_2=\frac{m_1}{m_2},\qquad
f_3=\frac{m_1^2+m_2^2}{m_2}\,;\\
g_1&=&2\,\frac{m_3}{m_2},\qquad g_2=\frac{n_3}{m_2},\qquad
g_3=-\,\frac2{m_2}\left( \left( m_1^2+m_2^2\right) m_3+m_1n_3\right).
\end{eqnarray}
Namely, we get
\begin{equation} \begin{array}{l}
2\,H_1 = S_1( f )+T_1( g ),\\
H_2 = S_2( f )+ T_2( g ),
\end{array} \end{equation}
where
\begin{eqnarray}
S_1 &=& \frac1{2}\left( (f_3+h_1f_1)^2 -4\,c_3f_1f_2-(c_4+h_2)f_1^2-
4\,h_1f_2^2\right) ,\\
S_2 &=& -\,2\,c_3(f_3+h_1f_1)f_2-(c_4+h_2)\,f_2^2-c_3^2f_1^2 ,\\
T_1 &=&\,\frac{1}{2}\,(-g_1g_3+g_2^2),\\
T_2 &=&
\frac{1}{4}\,((h_1\,g_1-g_3)^2-(c_4+h_2)\,g_1^2+4\,c_3\,g_1\,g_2).
\end{eqnarray}
From (20) it follows that the quantities $S_1( f )$ and $S_2( f )$
may be considered as a natural projection of two nontrivial
integrals of motion 2$H_1$ and $H_2$
\be \pi : 2 H_1 \to S_1(f);
\quad \pi: H_2 \to S_2(f). \ee
So we take them as the new dynamical variables .
It is natural also to unify them (as in (15)) to
\begin{equation}
{\cal F}(s, f ) = s^2 - S_1( f )\,s +S_2( f )
\end{equation}
and consider the equation
\be {\cal F}(s, f )=0. \ee
The roots $s_1$ and $s_2$
of equation (27) are the famous Kovalevskaya variables
\be s_1+s_2=S_1,\quad s_1 s_2=S_2 .\ee

Note that as it was shown by Novikov and Veselov [9], the
variables $s_1$ and $s_2$ are Poisson commuting,
\begin{equation}
\left\{ s_1,s_2 \right\}=0,\qquad \left\{ S_1, S_2 \right\} =0.
\end{equation}
Note also another property of these variables
\be
\{T_1,T_2\}=0,\quad 2\{H_1,S_2\}=\{H_2,S_1\}.\ee
So we have also
one-parametric family of Poisson commuting variables
$S_1(\lambda)=S_1+2\lambda H_1$ and $S_2(\lambda)=S_2+\lambda H_2$
\be \{S_1(\lambda),S_2(\lambda)\}=0.\ee
Note that functions $f_j, g_k$ are not independent
but they satisfy the relations
\be
\begin{array}{l}
f_1f_3-f_2^2=1,\\
f_1g_3+2\,f_2g_2+f_3g_1=0.
\end{array} \ee
These relations  are standard for the cotangent bundle of
two-dimensional two-sheet hyperboloid. So after change of
variables we come to the dynamical system on two-dimensional
two-sheet hyperboloid.

Namely in terms of new variables, equations of motion (5) have the
form
\begin{eqnarray}
\dot f_1&=& \frac12\,(f_1\,g_2+f_2\,g_1),\nonumber \\
\dot f_2&=& -\,\frac12\,(f_1\,g_3+f_2\,g_2)=\frac14\,(-\,f_1\,g_3+f_3\,g_1),\\
\dot f_3&=&-\,\frac12\,(f_2\,g_3+f_3\,g_2)\nonumber
\end{eqnarray} and
\begin{eqnarray}
\dot g_1 &=& c_3f_1^2+h_1f_1f_2-f_2f_3,\nonumber \\
\dot g_2 &=&\frac12\,\gamma \,f_1^2+c_3f_1f_2+\frac12\,f_3^2,\\
\dot g_3 &=&-\,c_3\left( f_1f_3+2\,f_2^2\right) -h_1f_2f_3- \gamma
_4 \,f_1f_2, \nonumber \end{eqnarray}
where
\[ \gamma _4=c_4-k^2=c_4+h_2-h_1^2. \]

Note also useful equations.
\begin{eqnarray}
\ddot f_1 &=& \nu f_1+\frac12\,(h_1f_1+f_3),\\
\ddot f_2 &=& \nu f_2+(h_1f_2+\frac12\,c_3f_1),\\
\ddot f_3 &=& \nu f_3+\frac12\left( h_1f_3-2\,c_3f_2-(c_4-k^2)f_1\right) ,\\
\nu &=& h_1-S_1. \end{eqnarray}

One can show that the equations of motion for $f_1$, $f_2$, $f_3$
have the Lax form
\be \dot L=[L,M]\,, \ee
where
\be
L=\left( \begin{array}{rr} f_2 & f_1\\ -\,f_3 &
-\,f_2 \end{array} \right),\qquad M=\frac14\left(
\begin{array}{rr}
-g_2 & g_1\\ -g_3 & g_2\end{array} \right) . \ee

The equations for $g_1, g_2, g_3$ have the form
\be
\dot M=[L,N]\,, \ee
where
\be N=-\,\frac{1}{8}\left(
\begin{array}{cc} c_3f_1+2\,h_1f_2, & f_3+h_1f_1 \\ \gamma _4
f_1+2\,c_3f_2-h_1f_3, &-\,(c_3f_1+2\,h_1f_2) \end{array} \right) .
\ee

Let us consider now the Clebsch problem [14]
(see also [15]--[18]), i.e. the problem of motion of rigid body in
ideal fluid. The dynamical variables here are the components of
momenta $p_1, p_2, p_3$ and angular momenta $l_1, l_2, l_3$ and
for special case they satisfy also
the additional constraint $(l,p)=0$.

This system is Hamiltonian relative to the Poisson structure for
the Lie algebra $e(3)$ of motion of three-dimensional Euclidean
space
\begin{equation}
\{l_i,l_j\}=\varepsilon _{ijk}\,l_k,\qquad \{l_i,p_j\}=\varepsilon
_{ijk}\,p_k,\qquad \{p_i,p_j\}=0,
\end{equation}
where $\varepsilon _{ijk}$ is a standard totally skew-symmetric
tensor.

The Hamiltonian has the form
\begin{equation}
H=\frac{1}{2}\left( \sum _{j=1}^{3} l_j^2 + \sum _{j,k=1}^{3}
B_{jk} p_j p_k \right)
\end{equation}
where the quantities $B_{jk}$ are constants.

One can check that the equations of motion for this case
have the same form as the equations (33), (34) and that the second
equation in (32) is equivalent to the condition $(l,p)=0$.

Note that for diagonal matrix $B$ this problem was solved by Weber
[3] in terms of abelian functions of two variables.

\bigskip\noindent
{\bf 4.} The last step is to reduce our problem to the case of
diagonal matrix $B$. For this it is convenient to use the
important identity discovered by K\"otter [2]: \be -\,2s\,{\cal
F}(s)= Q_2^2(s)-P_3(s)\,Q_1(s), \ee where
\begin{eqnarray}
P_3(s)&=& s P_2(s) +c_4s-2\,c_3^2,\nonumber \\
P_2(s)&=& s^2-2\,h_1s+h_2,\quad P_5(s)=P_3(s)P_2(s),\nonumber \\
Q_1(s) &=& f_1^2s-2\,f_2^2,\\
Q_2(s) &=& s^2f_1-(f_3+h_1f_1)s-2\,c_3f_2. \nonumber
\end{eqnarray}

Let us introduce instead variables $f_j, g_k$ the variables $x_j,
y_k$ by the formulae\footnote
{Such kind formulae were introduced by Weierstrass and they are
very useful in the theory of abelian functions .}
\begin{eqnarray} x_j&=&\sqrt{(s_1-a_j)(s_2-a_j)},\\
y_j&=&\frac{x_k
x_l}{s_1-s_2}\left(\frac{\sqrt{P_5(s_1)}}{(s_1-a_k)(s_1-a_l)}-
\frac{\sqrt{P_5(s_2)}}{(s_2-a_k)(s_2-a_l)}\right) ,\end{eqnarray}
where $a_j$ is the root of the equation $P_3(s)=0$ and $\{j,k,l\}$
is the cyclic permutation of $\{1,2,3\}$.

From (45) -- (48) we get the expression for  $f_j$ in terms of
$x_k$
\begin{eqnarray}
f_1&=& -i\,\sum _{j=1}^3\frac{\sqrt{2a_j}}{P_3'(a_j)}\,x_j, \\
f_2&=&i\,\sum _{j=1}^3\,\frac{\sqrt{a_ka_l}}{P_3'(a_j)}\,x_j,\\
f_3 &=& -h_1 f_1 -2i\,\sum _{j=1}^3
\frac{\sqrt{2a_j}}{P_3'(a_j)}\, (a_1+a_2+a_3-a_j)\,x_j
\end{eqnarray}
and
\begin{eqnarray}
g_1&=& -i\,\sum _{j=1}^3\frac{\sqrt{2a_j}}{P_3'(a_j)}\,y_j, \\
g_2&=&i\,\sum _{j=1}^3\,\frac{\sqrt{a_ka_l}}{P_3'(a_j)}\,y_j,\\
g_3 &=& -h_1 g_1 -2i\,\sum _{j=1}^3
\frac{\sqrt{2a_j}}{P_3'(a_j)}\, (a_1+a_2+a_3-a_j)\,y_j .
\end{eqnarray}
Note also that
\begin{equation}
Q_2(a_j)=i\sqrt{2 a_j}\,x_j, \qquad j = 1,2,3 .
\end{equation}

One can check that after substitution of expressions (49)--(54)
for $f_j$ and $g_k$ into the equations of motion (33), (34) the
equations for variables $x_j$, $y_k$ become the equations for the
special Clebsch case ($(l,p)=0$) with diagonal $B$ matrix. So we
may use the Weber solution [3] given by the formulae

\begin{equation}
x_j=x_{j0}\,\frac{\theta _{j4}(u_1,u_2)}{\theta
_0(u_1,u_2)},\qquad y_j=y_{j0}\,\frac{\theta _j(u_1,u_2)}{\theta
_0(u_1,u_2)},\qquad j=1,2,3, \end{equation}
where $x_{j0}$ and $y_{j0}$ are constants,
$u_1$ and $u_2$ are linear functions of
$t$ and $\theta _0(u_1,u_2)$, $\theta _j(u_1,u_2)$, $\theta
_{j4}(u_1,u_2)$ are standard theta functions with half-integer
theta characteristics:
\begin{eqnarray}
\theta _1 &=& \theta \left[ \begin{array}{ll} 1&0\\ 1&1
\end{array} \right] (u_1,u_2),\qquad \theta _2 =\theta \left[
\begin{array}{ll} 0&1\\ 0&1
\end{array}\right] (u_1,u_2),\qquad \theta _3 =\theta \left[ \begin{array}
{ll} 1&1\\ 1&0\end{array} \right] (u_1,u_2), \nonumber \\
\theta _{14} &=& \theta \left[ \begin{array}{ll} 0&0\\ 0&1
\end{array}
\right] (u_1,u_2),\qquad \theta _{24} =\theta \left[ \begin{array}{ll} 1&1\\
1&1 \end{array}\right] (u_1,u_2),\qquad \theta _{34} =\theta
\left[
\begin{array}{ll} 0&1\\ 0&0\end{array} \right] (u_1,u_2), \nonumber \\
\theta _0 &=& \theta \left[ \begin{array}{ll} 0&0\\ 0&0
\end{array} \right] (u_1,u_2). \,\end{eqnarray}
These functions are
defined by the standard formulae
\begin{equation}
\theta \left[ \begin{array}{ll} \varepsilon _1&\varepsilon _2\\
\delta _1&\delta _2 \end{array} \right] (u_1,u_2)= \sum
_{n_j,n_k=-\infty }^{\infty }\exp \,\left\{i\pi\left[ \tau _{j
k}\left( n_j+\frac{\varepsilon _j}2\right) \left( n_k+\frac
{\varepsilon _k}2\right) +\left( n_j+\frac{\varepsilon _j}2\right)
\left( 2u_j+\delta _j\right) \right]\right\} ,
\end{equation}
where $\tau _{jk}$ is the period matrix related to the algebraic
curve \be y^2=P_5(x).\ee So, the formulae (18), (19), (49)--(54)
and (56)--(59) give the explicit solution for the Kovalevskaya
top. ( For more details see [1],[2] and [3]).

\bigskip \noindent {\bf 5.}
Similarly to the Weierstrass approach for geodesics on an
ellipsoid [19], we may obtain once more important identity
\be
{\cal F}(s){\cal H}(s)-{\cal G}^2 (s)=2 P_5(s) \ee
where
\begin{eqnarray}
{\cal F}(s) &=& s^2-S_1 s+S_2, \\
{\cal G}(s) & =& \dot{S}_1 s-\dot{S}_2, \\
{\cal H}(s) &=& 2 (s^3-b_1s^2+b_2 s-b_3),\\
P_5 &=& (s^3-2 h_1 s^2+(c_4+h_2) s-2 c_3^2)(s^2-2 h_1 s+h_2)
\end{eqnarray}
and
\begin{eqnarray}
b_1 &=&-S_1+4 h_1,\nonumber  \\
b_2 &=& S_1^2-S_2-4 h_1 S_1 +2 h_2+4\,h_1^2+c_4,\\
b_3 &=& -S_1^3+4 h_1 S_1^2 +2 S_1 S_2-4 h_1 S_2
-(2 h_2 + 4 h_1^2 + c_4) S_1
- \frac{1}{2}\dot{S}_1^2+4 h_1 h_2 +2 h_1c_4+2 c_3^2 .
\nonumber
\end{eqnarray}
From (60) it is easy to get the equations of motion in standard
Abel-Jacobi form \be \dot {s}_1=i (s_1-s_2)^{-1}\sqrt{2
P_5(s_1)},\qquad \dot {s}_2=i (s_2-s_1)^{-1}\sqrt{2 P_5(s_2)} \ee
and also the Lax representation with spectral parameter
$s$ in terms of 2 by 2 matrices
\be \dot{\cal L}(s )=[{\cal L}(s ),{\cal M}(s)], \ee
\be {\cal L}(s )=\left( \begin{array}{rr} {\cal G} &{\cal F}\\
-{\cal H}& -{\cal G}\end{array} \right) ,\ee
\be {\cal M}(s )={\cal F}^{-1}(s)\left( \begin{array}{rr} 0 & 0\\
{\cal C}& {\cal D}\end{array} \right) \ee where \be {\cal
C}=(s-2\, T_1) {\cal F}(s)-(1/2) {\cal H}(s),\quad {\cal D}=-{\cal
G}(s).\ee

\bigskip\noindent {\bf 6.}
I would like to conclude this paper by the conjecture that similar
results are valid also for the $n$-dimensional generalization of
the Kovalevskaya top given in [4].

\section*{Acknowledgments}

The main result of this note on the equivalence of the
Kovalevskaya system to the special case of the Clebsch system has
been obtained in 1983 during the preparation of the review on the
motion of the rigid body around the fixed point [17]. Later  I had
the possibility to discuss with Prof. Moser the Kovalevskaya
problem and other problems of classical mechanics. These
discussions had great influence on my point of view on the
problems of classical mechanics in general. In particular, Prof.
Moser emphasized the important role of factorization of the
Kovalevskaya polynomial $P_5(x)$ into polynomials $P_3(x)$ and
$P_2(x)$. The simple explanation of this fact is absent
unfortunately till now.

I am grateful also to V.Z. Enolsky, A.P. Veselov and referee for their
remarks.

Finally, I would like to thank the Department of Physics of the
University of Oviedo for the hospitality.

\end{document}